# Social Networks Analysis in Discovering the Narrative Structure of Literary Fiction[1]

A. JARYNOWSKI, S. BOLAND

In our paper we would like to make a cross-disciplinary leap and use the tools of network theory to understand and explore narrative structure in literary fiction, an approach that is still underestimated. However, the systems in fiction are sensitive to reader's subjectivity and attention must to be paid to different methods of extracting networks. The project aims at investigating into different ways social interactions are *read* in texts by comparing networks produced by automated algorithms-natural language processing (NLP) with those created by surveying more subjective human responses. Conversation networks from fiction have been already extracted by scientists, but the more general framework surrounding these interactions was missing.

We propose several NLP methods for detecting interactions and test them against a range of human perceptions. In doing so, we have pointed to some limitations of using network analysis to test literary theory (e.g. interaction, which corresponds to the plot, does not form climax).

## Introduction and background information

A social network is the map of interactions between individuals. We use social network analysis to understand and explore social structures, such as whether the friend of my friend is also my friend – or whether the enemy of my friend is also my enemy. Social network analysis has been applied to human societies by sociologists and to animal societies by biologists for many years. In this project, we want to make a cross- disciplinary leap, and use the tools of social network analysis to explore social structure within fiction. We could investigate such questions as: Can we characterise the genre or period of a book from aspects of the social network it represents? More specifically, do the social networks described by authors writing about well-functioning societies differ in recognisable ways from those writing about dystopia? Does social network analysis provides ways to articulate plot developments in novels? To what extent does social network analysis correlate with the interpretative activities of different readers, for example, do different readers infer different social networks from the same novel? The results of this project will provide new interdisciplinary insights, and will demonstrate to what extent social network analysis can be usefully applied to the study of narrative.
Here we analyze Sherwood Anderson's short story series *'Winesburg, Ohio'*. We compare networks produced by automated systems with those created by surveys (more subjective human responses). Exploration of the biographical and genetic (i.e. structural) process behind the writing of

1 We thank Elva Robinson, Dan Franks, John Forrester, Richard Walsh from YCCSA at York Univ.

Winesburg, Ohio allows for dealing with the social networks [1], because it reveals something of how character relationships grew. We situate our project in relation to this story to offer a literary perspective (narrative theory with a rhetorical-pragmatic bias). We want to observe the tension between narratives and complex systems in various interdisciplinary manifestations [2].

The project aims to investigate various ways social interactions 'read' in texts. We have investigated different ways of understanding interaction in literary fiction. Quantitative studies have been used in literary studies or linguistics for very long time. The best example of this approach is analysis using Zipf [3], which studies the frequency distribution of words in text. It turns out that by observing the frequency of words, writers and literary genres can be distinguished. In addition, many researchers are analyzing additional connections between words (their location in a sentence) that can be represented as a network [4]. In such a network the nodes (words) are linked by lexico-semantic relations. We also study such networks, but our object of analysis are characters.

In this article, network analysis is a basic methodological tool. We began by examining the theory of complex networks (dynamically developing science) via random graphs [5, 6]. An important advantage of the network approach is its vast field of its potential applications. From physical systems to biological and social, you can use them virtually anywhere where there are dependencies between elements. The border between science and society has given rise to a very popular research technique: social network analysis (called Social Network Analysis - SNA). The structure of relationships between people has a significant impact on many factors, such as the flow of information, or the ratio of power. The position of the individual in the network also determines many of the features, such as social position. As part of the theory of SNA we distinguish between binary (connection between components exist or not) and weighted (connections may have different weights, reflecting different levels of intensity of the interaction) variables. In our case, the intensity of the relationship will be important in determining a more accurate description of the community which is of interest to us. There are also an additional distinction between directed networks (connection between components has a fixed direction) and undirected (connections do not have a direction, and recorded only the relationship of reciprocity).

The fields of NLP (Natural Language Processing) and Data Mining have already developed set of tools in Text Processing. Researchers in science and commerce have used them for a long time [7]. New approach have been developed at Stanford [8] and Columbia [9] for literary fiction. Both tried to extract networks from literary fiction. From a computational point of view, Stanford's research [8] was very simple. They analysed dramas and define an interaction as the coincidence of two characters in once scene. This simple method allowed them to extract networks and test them against hypotheses from literary theory. On the other hand, Columbia looked at conversations in prose texts and defines speech between characters as an "interaction". This paper [9] is a useful starting point for considering further possibilities. However, this kind of approach stands in need of critique for a range of reasons, and we will discuss some perspectives later in our paper.

## Preliminary analysis

### *Interactions as communication and distinguishing types of relation*

Early on in the project, several ways of dividing social interactions were posited: interactions that were 'implicit' and 'explicit', interactions that were primarily 'thematic' or 'plot-driven', etc. It was decided that the only distinction that was to be made in the surveys was between interactions that are 'important' and 'unimportant', with the reader left to decide what they considered important ('your reading of the story'). This was then further developed to allow the reader the ability to rate each interaction's importance out of ten, giving, hopefully, a more nuanced result.

Attention must be paid during later analysis to the difference between actions important within the world of the story, and actions that form an important part of the reading experience. This distinction must be considered during both interpreting the human survey results and in developing a method by which to automatically 'read'.

Certain limitations had to be imposed. We started from theory of information and used a schema — sender – message - receiver — which implies direction in channel of communication. Sometimes it seems to be difficult to describe precisely who is who in this construction but is possible (more or less objectively). We kept in mind that a “message” is not only a speech or letter, but also nonverbal and even non-physical forms of communication which can take place in the story. In those nontrivial events, let us assume that the sender is the subject who impacts on the receiver, the object.

The first network [Fig.1] were generated by reading of the short story 'The Teacher’. In this context, an 'explicit' interaction is one which is shown in the text, i.e. two people speaking or touching. An 'implicit' action would perhaps be better described as an 'indirect' one; it is something referred to or remembered, rather than shown clearly to the reader. Although this distinction permits an interesting investigation into whether or not interactions we 'see' are judged as more important, it was dropped from the full pilot survey due to ambiguity.

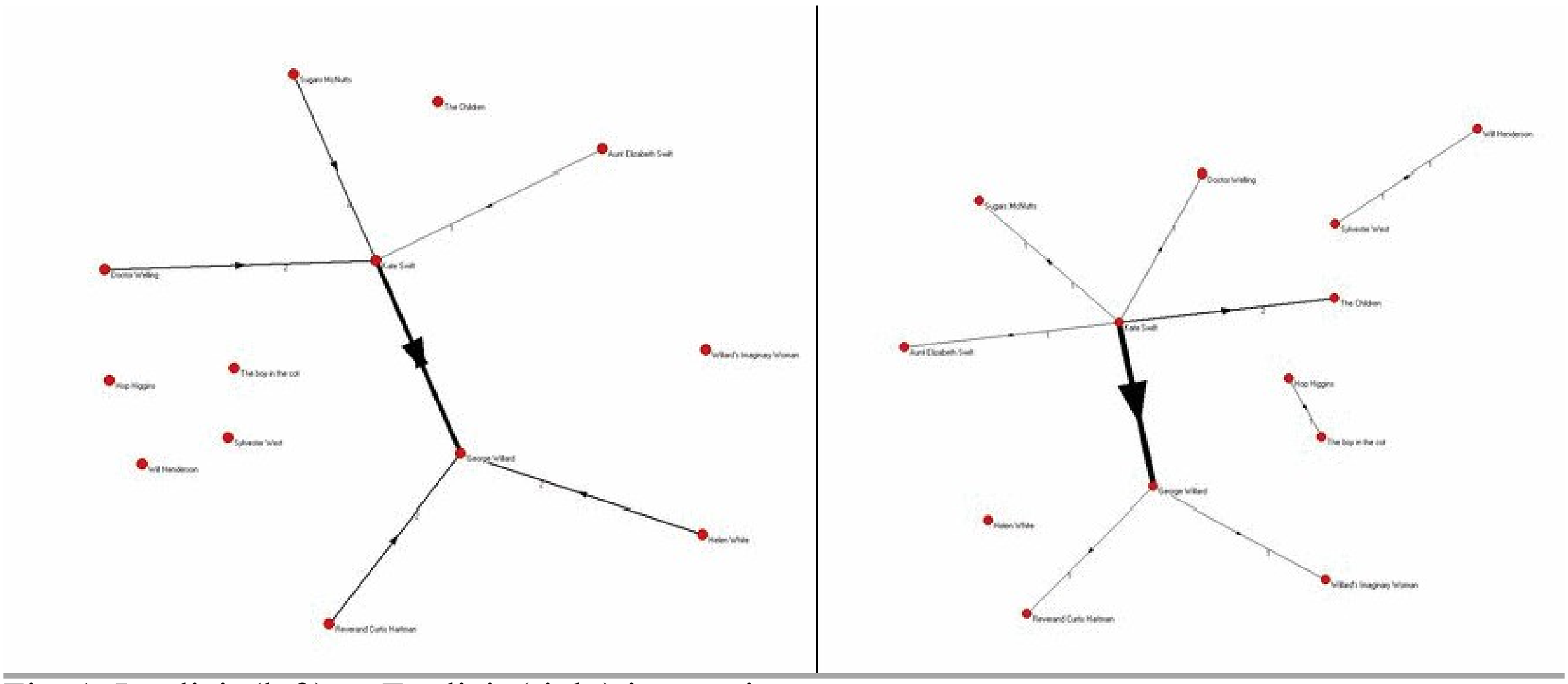

Fig. 1. Implicit (left) vs Explicit (right) interactions

### *Interactions undistinguished, but significance added, as well as overall perception*

Models [Fig. 2] show the difference between interactions noted (equivalent of Task 1) with annotated significance in scale:  'YES', 'NO' while going through the text, and the overall importance (equivalent of Task 2) of the relationships upon the story. Students and supervisors took part in this pre-pilot. In this version, one network is directed and second not. Importance is measured on different scales: once binary ('YES', 'NO') second linear (0, 1, 2, .., 10). Datasets of both parts also differ. In the annotative survey  ('quasi' Task 1) the same interaction can appear few times, but in overall view (Task 2) respondent is allowed to entry once every element of half-matrix (where row and column indicate who interact with whom).

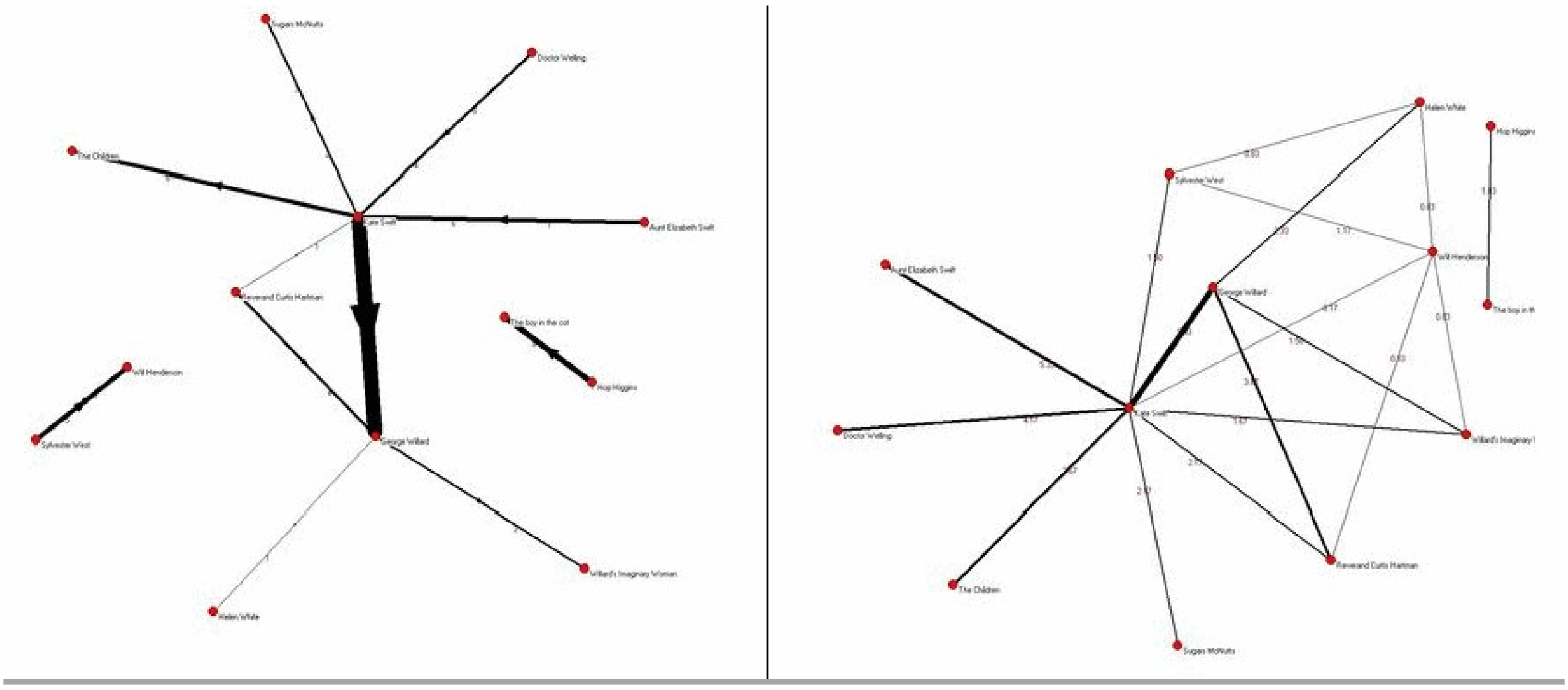

Fig. 2. Interactions noticed (left), Overall view (right)

### *Interactions undistinguished, but significance added*

As the final version, we forget about communication schema and finish with undirected networks (the same types computer-detection could only give to us). Both parts: noticed (Task 1), overall (Task 2) have the same scale (0-10).

## Data collection technical assessment

### *Computer-detection*

In our project, we focus on a more 'human' method of detecting interactions. As we explained earlier, interaction is sometimes more than conversation: it can be physical, or even mental (one character thinking of another) etc. We find coincidence in unit of text (paragraph, sentence) as a best indicator of wide-defined interaction between characters. A much more difficult task was to come across a method for determining the importance of an association. This is a new element not analysed in existing papers in the field. The only relevant approach we could imagine is biased without that comparison with human could be broken. Importance is obtained via multiplication of frequencies of character's names appearing in a unit. Another simplification came from recognition of characters. To do that a list of potential aliases was built by us.

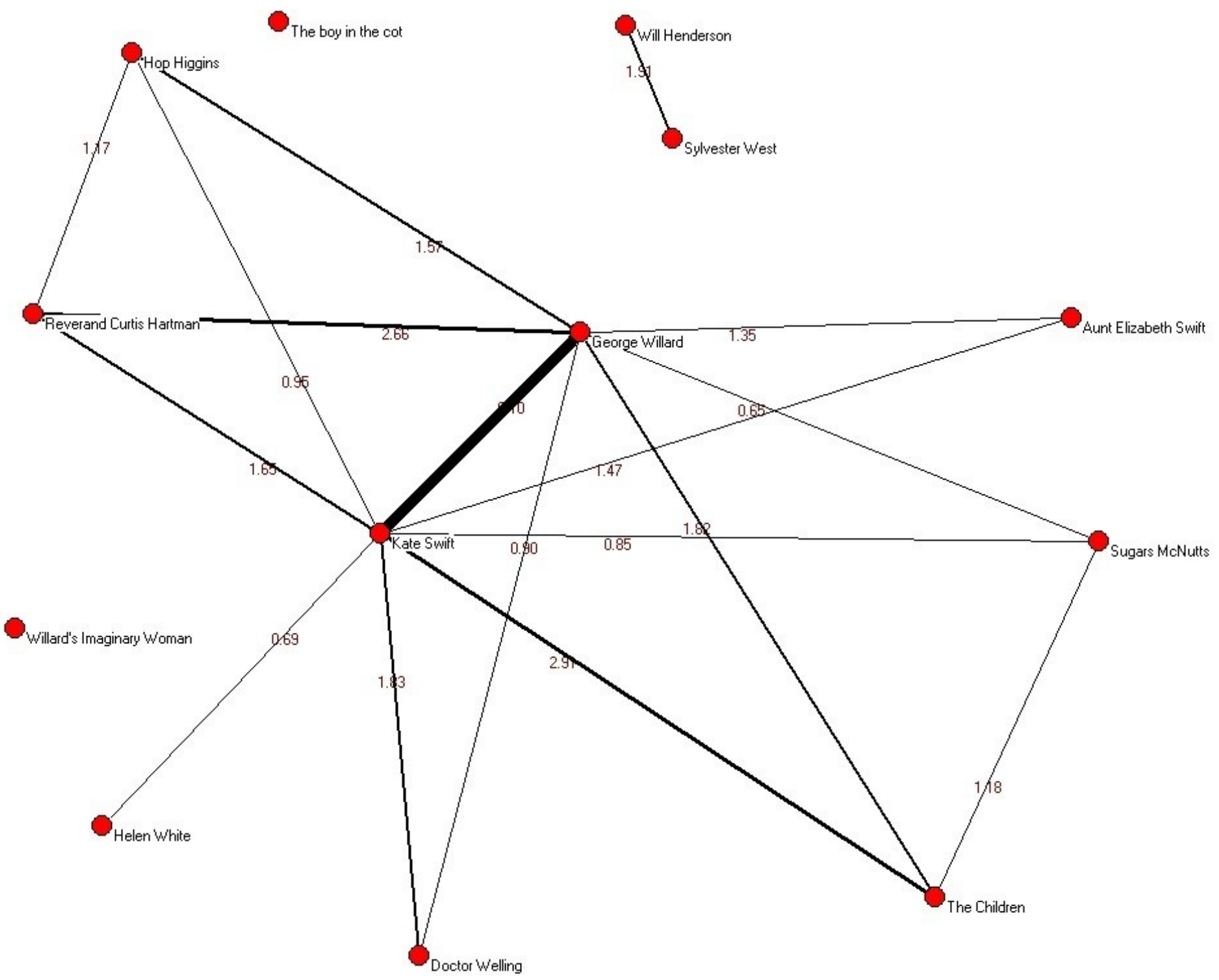

Fig. 3. Example of computer-detection network with paragraph as a unit

All computation of automatic detection was made in RapidMiner with R plug-in.

***PHP- survey specification***

Survey was set up using PHP and data were collected in MySQL server provided by Computer Science Department (Univ. of York). Respondents were asked to fill in 4 pages: a consent form, task 1, task 2, and metrics. The server remembered each respondent by an individual id based on IP address (there was a possibility for the survey to be filled in by more than one person using the same IP address). Analysis was anonymous but respondents had a chance to leave us their email address if they wished to be contacted with the results. The server was constructing to be used with different stories and list of characters has been imported from respective table (we ran it for two stories: “The Teacher” and ”The Philosopher”).

# Data Analysis

There are three methods [Fig.1], two of them are human based. In Task 1, respondents were asked to enter every interaction they recognized with and give it a value of importance using a scale [1-10]. In Task 2, respondent entry lower-matrix, which represent links between characters also in scale [1-10] where empty entries give 0 in our scale – meaning no link at all. A normalization problem appears, because in the first task links can be represented by more than entries and the weight of link is a sum of all entries. This brings about a situation in which some links can have value above 10 (which is a limit in Task 2). On the other hand, the computer counts multiplication of frequencies of character's alias appearing in each unit and add 1. That bring scale [1- …]. As in Task 1, final link weight is a sum of all entries for related characters. We observed that scale of Task 2 is the most efficient, so we decided to use this scale as a pattern. Our idea was: let the most significant link have the same value for those 3 networks [Fig. 2]. Then the rest of the links have to be changed in respect to the selected one. To apply this rule in computer task  sum of multiplications were corrected with linear factor, that final most common link has the same value. The same rule was applied in Task 1.

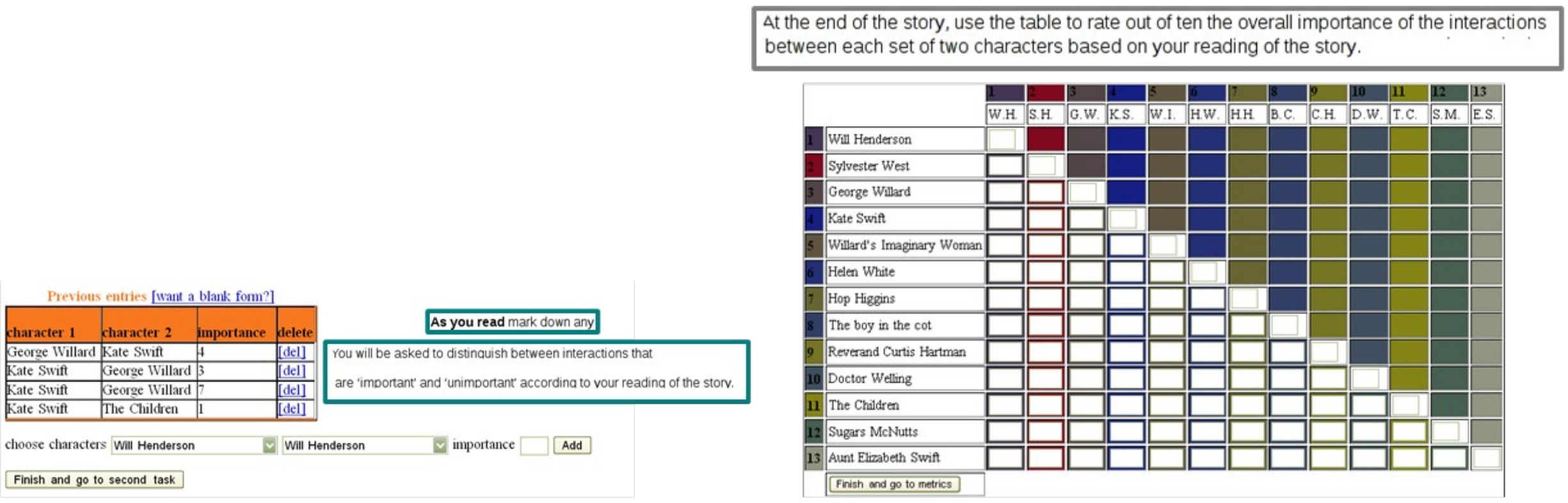

Pic. 1. Picture of 2 methods of inputs in 2 different scales (left: 1 Task-entries, right: 2 Task-matrix)

### *Normalization of perspectives*

We observed, different respondents had different patterns of filling survey [Fig. 3].

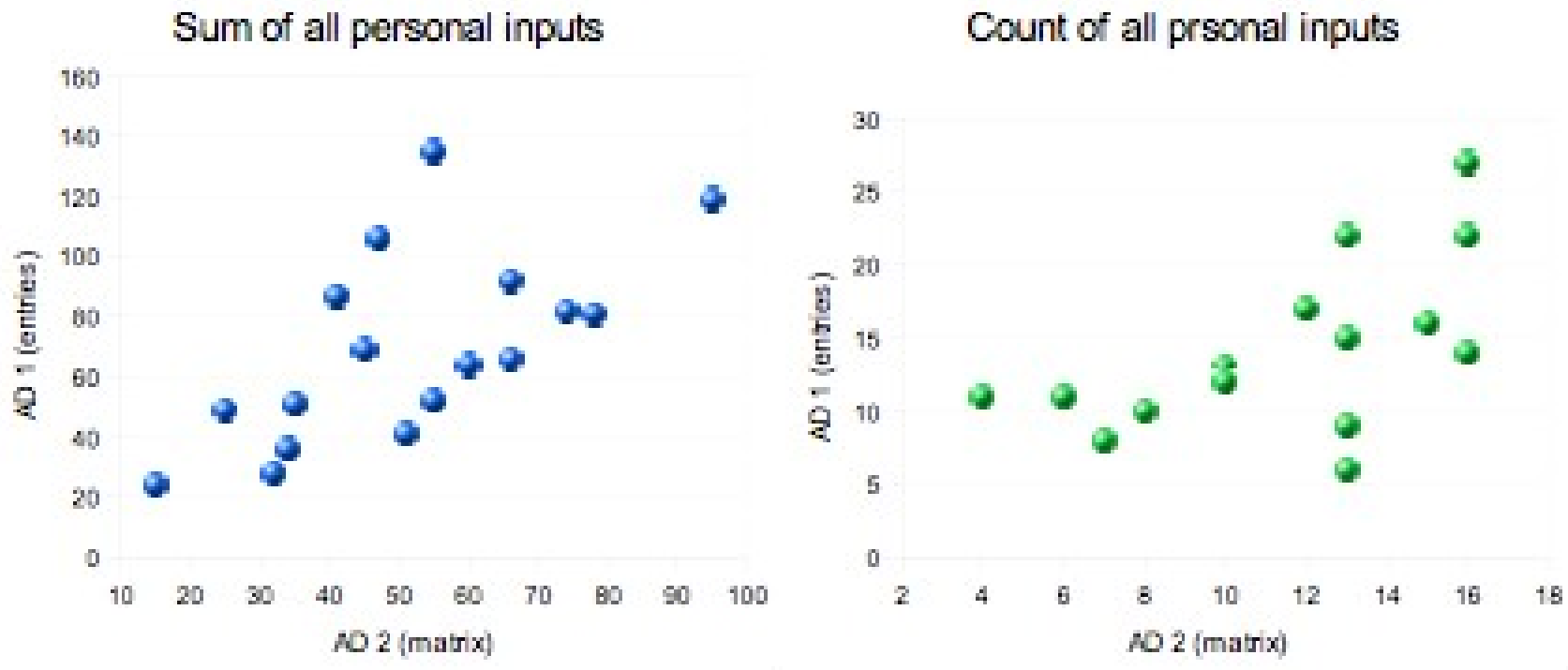

Fig. 4. Numbers of entries by different respondents (17 of them).

To cope this that, let every respondent have the same value (democracy). All respondents inputs are normalized to the average response. Of course, this rule is applied only to the comparative analysis between methods and not related to respondent statistics, where the pattern chosen can tell us how each respondent reads the text.

One of the side projects was to investigated methods of reading. On [Fig. 4] there are correlations between Task 1 and 2 between first cohort of respondents. Some correlation are very low in full study [Tab. 2]. Those extreme value come from the fact that those respondents put almost every possible “interaction” in the story into their Task 2 entries and assigned most of them a value of 1, likely believing that every character has some connection with the others, but most of them are insignificant.

# Results

### *Networks*

From collected information from people, we can obtain average networks [Fig. 5,6]. Those networks are weighted (by importance mentioned by human or calculated by computer), but we are analyzing also binary networks, where links exist or no.

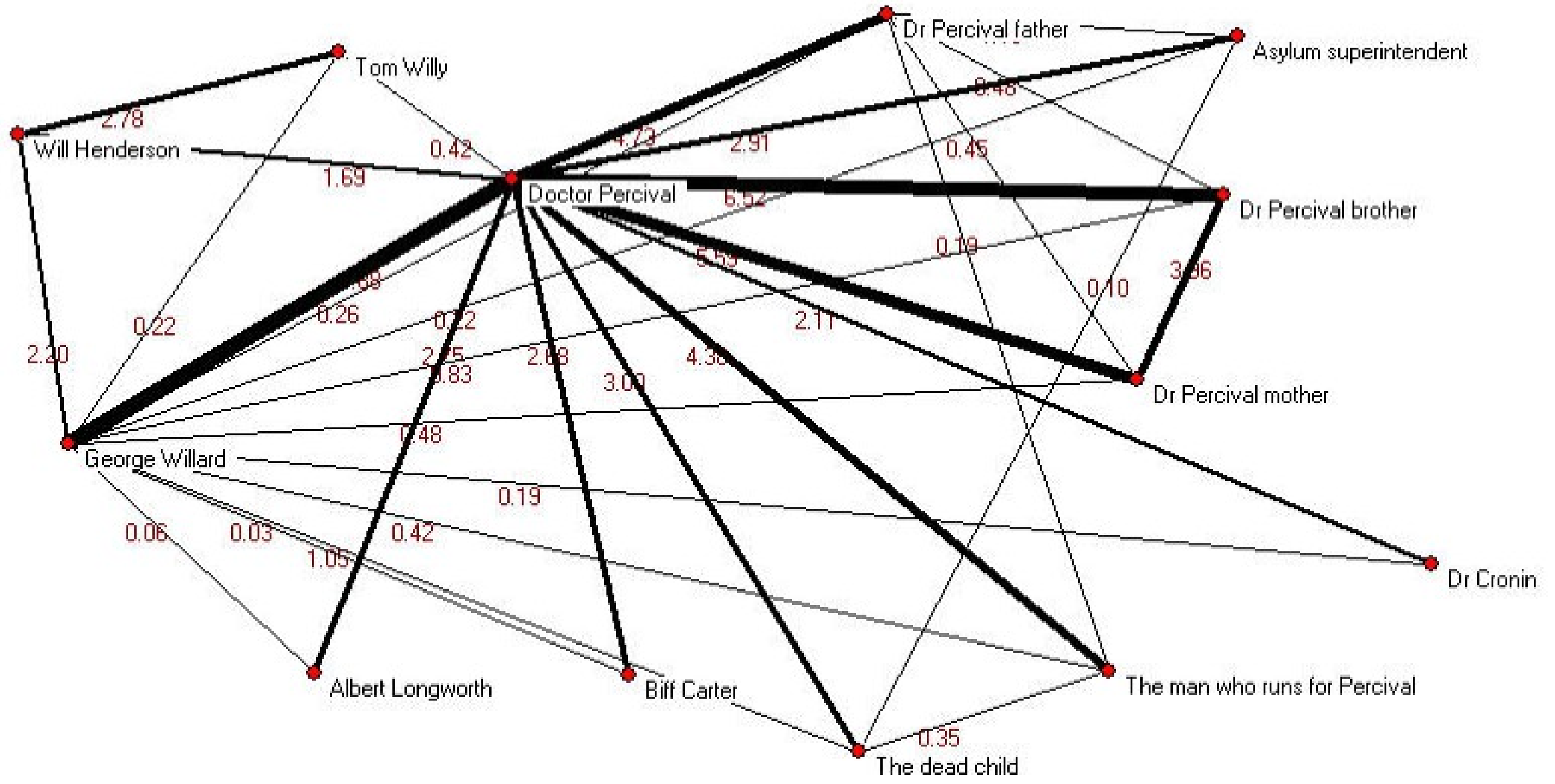

Fig 5. Final human-based network of Task 2 (matrix-global)

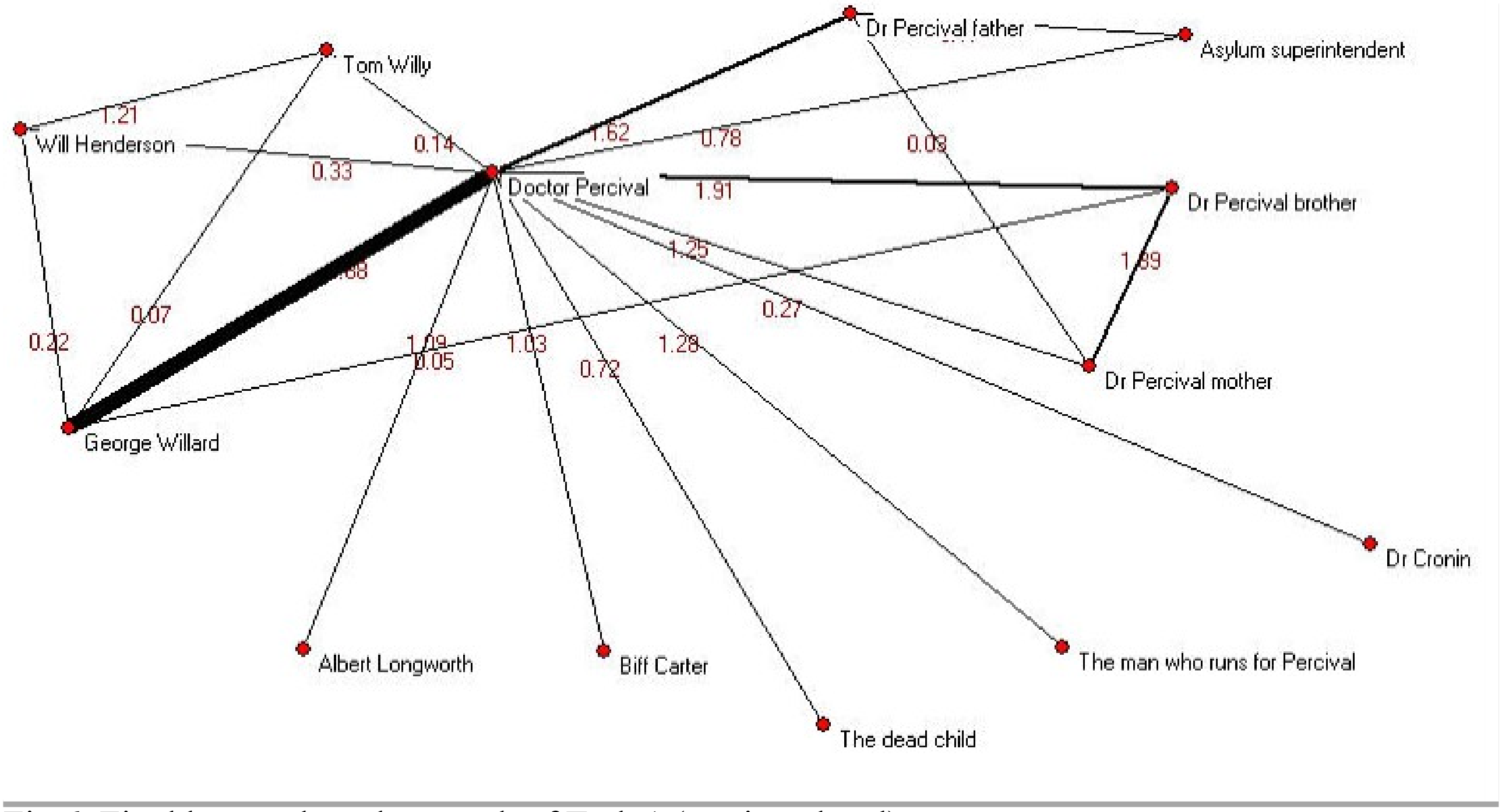

Fig 6. Final human-based network of Task 1 (entries - local)

We can also compare [Tab. 1] it with computer network [Fig. 7]

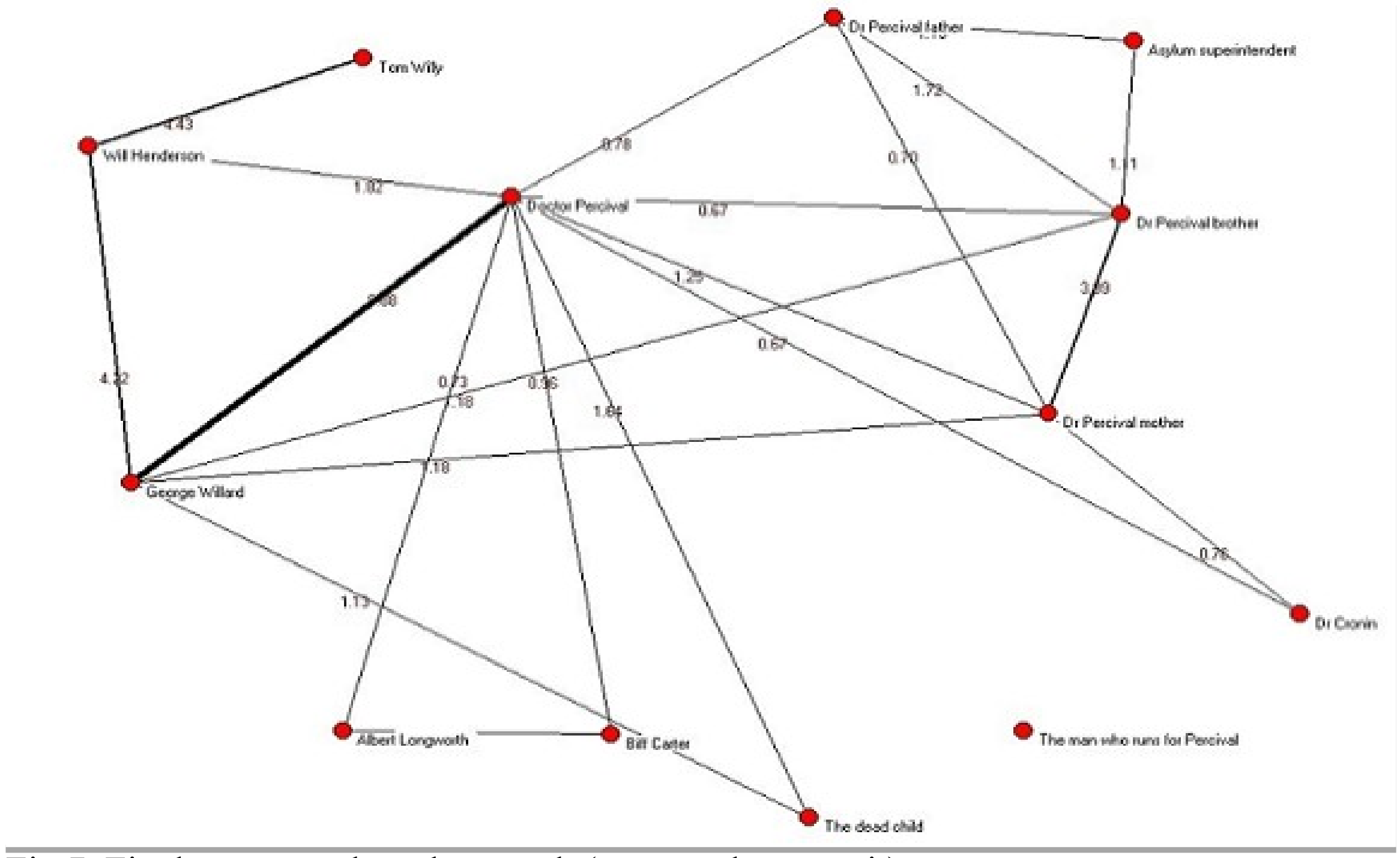

Fig 7. Final computer-based network (paragraph as a unit)

| Task 1 | Task 2 | Computer |
|---|---|---|
| ------------------------------------- | ------------------------------------- | ------------------------------------- |
| Arcs | Arcs | Arcs |
| ------------------------------------- | ------------------------------------- | ------------------------------------- |
| Number of lines with value=16 | Number of lines with value=24 | Number of lines with value=21 |
| ------------------------------------- | ------------------------------------- | ------------------------------------- |
| Density = 0.1025641<br>Average Degree = 2.4615385 | Density = 0.1538462<br>Average Degree = 3.6923077 | Density = 0.1346154<br>Average Degree = 3.2307692 |

Tab. 1. Comparison of networks properties between different methods

### *Correction of probably false links.*

Task 2 gave some links which do not appear in Task 1. As researchers, we know that some of them couldn't be found in any reasonable way. Our idea for correction is thus: If a value is far away from the mean then assume it should be 0 (there is no 'real' link). This value is probably due to human error. We propose two ways of correction, 2 and 3 sigma (well known from statistics).

- Correction 1 (3sigma): Delete link if distance from mean is greater than 3 sigma (standard deviations). This algorithm delete extreme outliers only.
- Correction 2 (2sigma): Delete link if distance from mean is greater than 2 sigma (standard deviation). This algorithm delete all suspected errors, but also some potential in-directed relations.

Described approach was invented for Task 2, because respondents could have problems with, for instance, entering data in the wrong cell or confusing the names of characters.

Another method can be applied to binary networks. Let us assume that same fake links disappear if we set up some threshold for links. In our case 11 seems to be ideal threshold, because it was found that at least two respondents found this link (maximal importance is 10).

### *Correlation between methods*

The final stage of our project is comparable analysis of various human and computer approaches. The overall view is presented in correlation matrixes. Every cell represents a correlation coefficient [10] between rows and columns [Tab. 2], calculated by comparing 2 weighted networks (matrixes). If we want to conclude results (all correlations are significant), let us start from computer methods. Almost everywhere, the paragraph method seems to be better (more comparable to human) than the sentence method. Opposite relation is only for weighed networks of Task 1. That can be explained by the method of reading and detecting interactions in this task: respondents could not enter an interaction if the character name appeared in next or previous sentence from that in which the interaction appeared. The same as computer. If we look at correlation between Task 1 and Task 2 with correction 2, we see it is literary 1 for weighted networks, so there are identical.

| | Paragraph – based algorithm | Sentence – based algorithm |
|---|---|---|
| Task 1 (entries) | 0.84 | 0.91 |
| Task 2 (matrix) | 0.70 | 0.58 |

Tab. 2) Correlations between Task 1 and 2 for whole population

We also did a statistical test for a few factors which can explain this coefficient. We choose some demographic factors such as gender, age, educational level, academic background, which are qualitative. On the other hand there were some quantitative factors like sum of entries for both tasks. To do so logistic regression was used [Tab. 3].

| Academic background | Parameter | p-Value |
|---|---|---|
| Arts and humanities | 0.05 | 0.37 |
| Social Science | – 0.03 | 0.71 |
| Science and medical Science | 0.16 | < 0.01 |

Tab. 3) Logistic regression for correlation between tasks.

These correlations be explained by:
- Demographic factors - only background is statistically significant. If respondent come from science, correlation seems to be better;
- The survey process - the sum of weight given in Task 2 is statistically significant and relation is negative. If the respondent enter many links, the correlation seems to be lower.

### *Climax formation*

The last point of analysis is testing a hypothesis of networks in literature. Here, we attempt to build a climax function [Fig. 8], constructed by networks.

The idea is to examine intensity and importance of interactions, as it corresponds to the basic plot structure offered by Freytag's Triangle. The sum of weighted interactions indicate narrative importance. We test to see if this climax curve can be produced automatically, using a quantitative method to seek qualitative results.

Again, our study compared human (Task 1) and computer method (both consist time order structure), with the story divided into four sections.

- Respondent: Human's input, divided into 4 parts, in time order: "Importance" is naturally indicated by respondents.

- Computer: Text divided into 4 parts (containing paragraphs) by numbers of words. This algorithm delete all suspected errors, but also some potential in-directed relations.

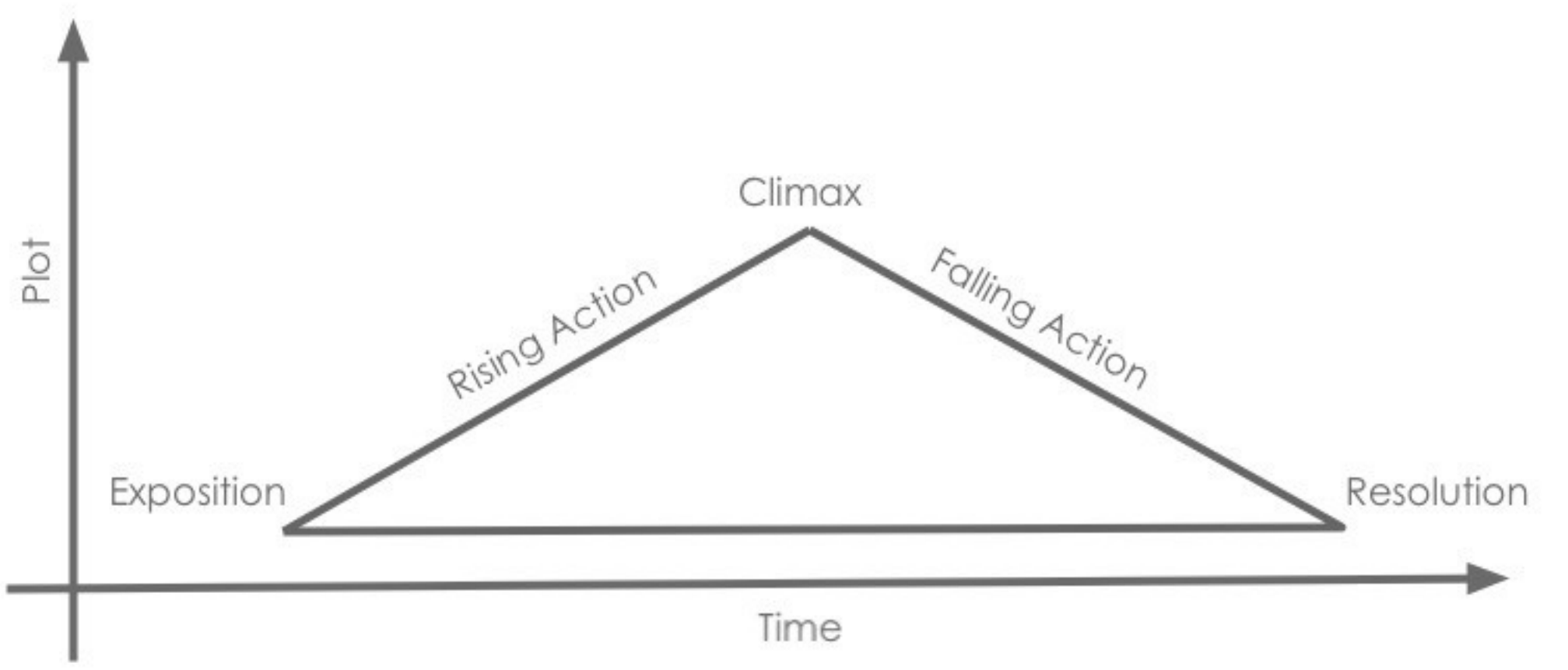

Fig. 8) Climax formation [11]

We create graph from data [Fig. 9] and we found anti-climax formation. We can interpret that in some ways. Firstly, computer algorithms found so many interactions between characters in the beginning of plot, because the author must introduce them. On the other hand, many of our human respondents put the most significant interactions at the end of the story, leading to a large input there.

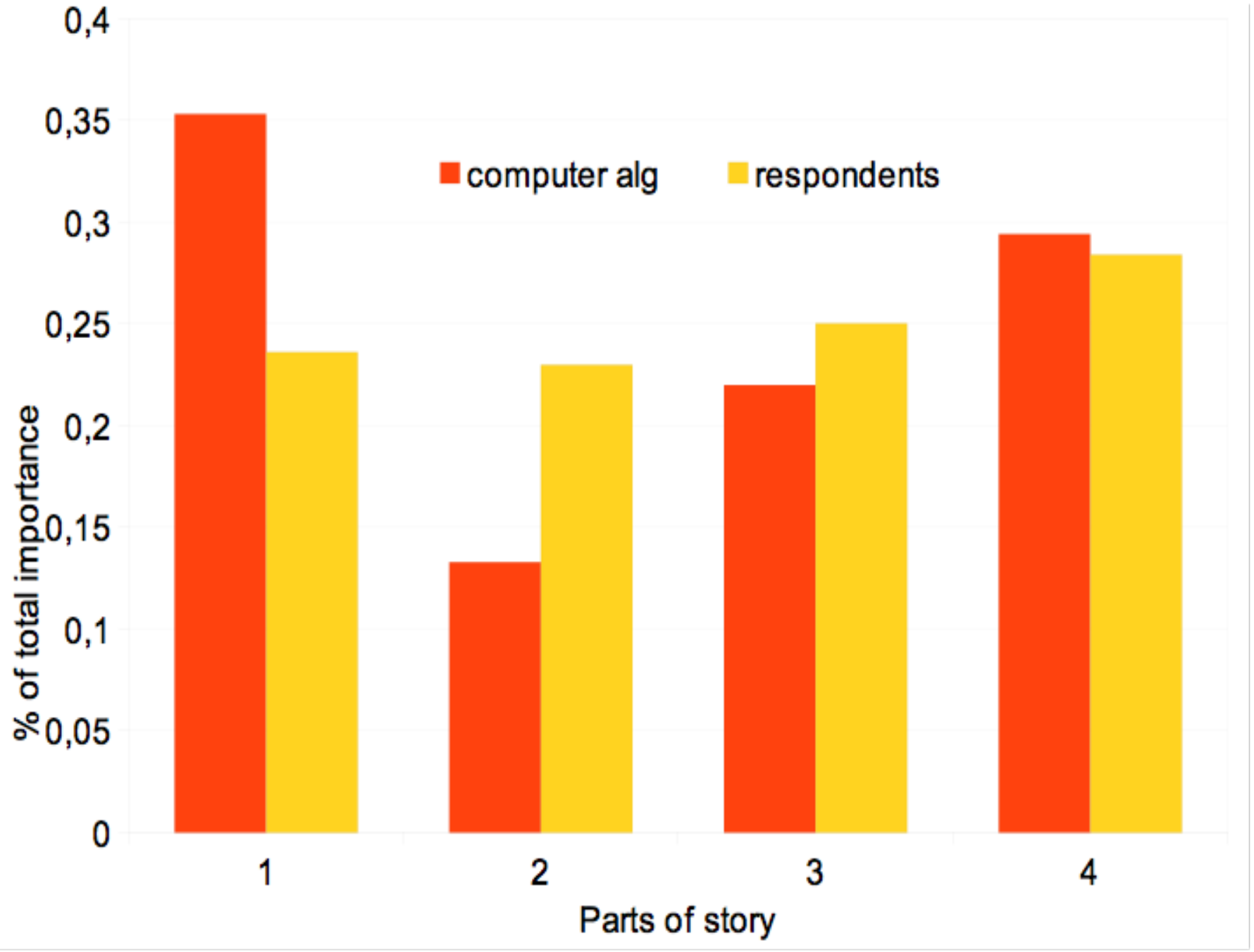

Fig. 9) Anti-climax formation from normalized data

# Conclusions

Discovering social networks in the literature turned out to be a fascinating subject but, it is still a pioneering area of research. We have developed methods of obtaining network based on surveys from readers of *Winesburg, Ohio* (surveys can be used to analyze novels). In our research we have simplified way of gathering information (from reader) in order to preserve the possibility of comparison with the computer methods. For example, we do not put the respondent in any particular readership context. Even preliminary analysis showed that the way of asking questions is important. We focus on reader subjective perspective (reader world).

The most important result seems to be correlations between the answers of questionnaires and algorithms (Tab. 2). Different level of correlation between the respective computer methods and different readers tasks shows the ambivalence and sensitivity to external factors. No universal method is available, which was a paradigm of NLP. For communication entries (task 1) view of networks shorter text window algorithms work better, but for general perspective (task 2), longer text window should be applied. There is also interesting statistically significant relationship between the perception of the network and education of respondent (people of science build the network more reproducible than the rest, potentially due to the different forms of reading encouraged by literary syllabi during school versus those encouraged by a humanities degree). Our research therefore not only explored ways in which a computer might be able to 'read' aspects of narrative, but also raised the question of how closely different readers' interpretations adhere to certain objective signifiers in the text.

Ultimately, our results suggest the intensity of interaction does not directly correlate with subjective reader interpretation of when tension builds in the plot. Concluding, the described methodology can assist in verifying hypotheses relating to specific texts, authors or generally – literature, while network based NLP tools getting more and more popular [12] and have potential to further research for digital humanities.